\documentclass[aps,prl,superscriptaddress]{revtex4-2}

\usepackage{graphicx}
\usepackage{amsmath}
\usepackage{upgreek}

\usepackage{color}
\usepackage{hyperref}
\begin{document}

\title{Two-dimensional optomechanical crystal resonator in gallium arsenide}

\author{Rhys G. Povey}
\email{rhyspovey@uchicago.edu}
\affiliation{Department of Physics, University of Chicago, Chicago IL 60637, USA}
\affiliation{Pritzker School of Molecular Engineering, University of Chicago, Chicago IL 60637, USA}

\author{Ming-Han Chou}
\affiliation{Department of Physics, University of Chicago, Chicago IL 60637, USA}
\affiliation{Pritzker School of Molecular Engineering, University of Chicago, Chicago IL 60637, USA}

\author{Gustav Andersson}
\affiliation{Pritzker School of Molecular Engineering, University of Chicago, Chicago IL 60637, USA}

\author{Christopher R. Conner}
\affiliation{Pritzker School of Molecular Engineering, University of Chicago, Chicago IL 60637, USA}

\author{Joel Grebel}
\affiliation{Pritzker School of Molecular Engineering, University of Chicago, Chicago IL 60637, USA}

\author{Yash J. Joshi}
\affiliation{Pritzker School of Molecular Engineering, University of Chicago, Chicago IL 60637, USA}

\author{Jacob M. Miller}
\affiliation{Department of Physics, University of Chicago, Chicago IL 60637, USA}
\affiliation{Pritzker School of Molecular Engineering, University of Chicago, Chicago IL 60637, USA}

\author{Hong Qiao}
\affiliation{Pritzker School of Molecular Engineering, University of Chicago, Chicago IL 60637, USA}

\author{Xuntao Wu}
\affiliation{Pritzker School of Molecular Engineering, University of Chicago, Chicago IL 60637, USA}

\author{Haoxiong Yan}
\affiliation{Pritzker School of Molecular Engineering, University of Chicago, Chicago IL 60637, USA}

\author{Andrew N. Cleland}
\email{anc@uchicago.edu}
\affiliation{Pritzker School of Molecular Engineering, University of Chicago, Chicago IL 60637, USA}
\affiliation{Center for Molecular Engineering and Materials Science Division, Argonne National Laboratory, Lemont IL 60439, USA}

\date{\today}

\begin{abstract}
In the field of quantum computation and communication there is a compelling need for quantum-coherent frequency conversion between microwave electronics and infra-red optics. A promising platform for this is an optomechanical crystal resonator that uses simultaneous photonic and phononic crystals to create a co-localized cavity coupling an electromagnetic mode to an acoustic mode, which then via electromechanical interactions can undergo direct transduction to electronics.
The majority of work in this area has been on one-dimensional nanobeam resonators which provide strong optomechanical couplings but, due to their geometry, suffer from an inability to dissipate heat produced by the laser pumping required for operation.
Recently, a quasi-two-dimensional optomechanical crystal cavity was developed in silicon exhibiting similarly strong coupling with better thermalization, but at a mechanical frequency above optimal qubit operating frequencies.
Here we adapt this design to gallium arsenide, a natural thin-film single-crystal piezoelectric that can incorporate electromechanical interactions, obtaining a mechanical resonant mode at $f_\mathrm{m} \approx 4.5\,\mathrm{GHz}$ ideal for superconducting qubits, and demonstrating optomechanical coupling $g_\mathrm{om}/(2\,\pi) \approx 650\,\mathrm{kHz}$.
\end{abstract}

\maketitle

\section{Introduction}
Spurred on by the promise of quantum computing \cite{Steane_1998,Preskill_2018},
superconducting qubits \cite{Nakamura_1999,You_2003,Blais_2004,Frunzio_2005,Wendin_2017,Kwon_2021,Gao_2021}
have become a ubiquitous topic of research and development.
These devices, typically operating with microwave electronic frequencies of $4\text{--}8\,\mathrm{GHz}$ and housed in dilution refrigerators at $\sim10\,\mathrm{mK}$, are excellent candidates for quantum information processing but suffer from an inability to engage in long-range communication due to the lossy environment and large amount of microwave thermal noise at room temperature.
To facilitate such connections outside of cryogenic temperatures, a quantum transducer that is able to coherently convert a microwave-electronic quantum state to an infra-red fiber optic signal, ideally in the telecom band of $1530\text{--}1565\,\mathrm{nm}$, is desired \cite{Lambert_2020,Lauk_2020,Chu2020,Moody_2022}.
Of the many approaches being explored for this challenging task, a propitious route is through an optomechanical crystal cavity \cite{EichenfieldOMCbeam,SafaviNaeiniOMCdesign} that emerges by placing a defect in a simultaneous photonic and phononic crystal. This defect site is able to support both a microwave-frequency acoustic resonant mode and an infra-red electromagnetic resonant mode, allowing them to couple to each other. Acoustic modes can then be directly transduced to electrical signals through a piezoelectric material.

The optomechanical Hamiltonian for an optical resonant mode at $\omega_\mathrm{o}$, annihilation operator $\hat{a}$, and a mechanical resonant mode at $\omega_\mathrm{m}$, annihilation operator $\hat{b}$, is given by
\begin{equation}
\hat{H} = \hbar\,\omega_\mathrm{o}\,\hat{a}^\dagger\,\hat{a} + \hbar\,\omega_\mathrm{m}\,\hat{b}^\dagger\,\hat{b} +  \hbar\,g_\mathrm{om}\,\hat{a}^\dagger\,\hat{a}\,\bigl(\hat{b}^\dagger+\hat{b}\bigr)\;,
\end{equation}
where the interaction term, with optomechanical coupling $g_\mathrm{om}$, can be derived by considering an optical Fabry-P\'{e}rot mirror attached to a mechanical spring undergoing small oscillations \cite{OMCavityReview,CavityOptomechBook}.
To obtain an exchange interaction we need to include a strong laser pump at frequency $\omega_\mathrm{L}$,
adding the term $\hbar\,L\,\hat{a}^\dagger\,\mathrm{e}^{-\mathrm{i}\,\omega_\mathrm{L}\,t}$,
where $L$ is measure of the pump strength.
For a strong pump we can expand $\hat{a}(t) = \hat{a}^\prime(t) + \bar{a}\,\mathrm{e}^{-\mathrm{i}\,\omega_\mathrm{L}\,t}$, with $\bar{a} \propto L$, and linearize for $\bar{a} \gg \hat{a}^\prime$.
With the laser red-detuned from the optical resonance, $\omega_\mathrm{L} = \omega_\mathrm{o} - \omega_\mathrm{m}$,
under the rotating wave approximation our interaction term then becomes
\begin{equation}
\hat{H}_\mathrm{int} \approx \hbar\,g_\mathrm{om}\,\bigl(\bar{a}^\ast\,\hat{a}^\prime\,\hat{b}^\dagger+\bar{a}\,\hat{a}^{\prime\,\dagger}\,\hat{b}\bigr)\;, \label{eq:OMintswap}
\end{equation}
providing a `beam-splitter' swapping interaction that allows infra-red optical excitations to be exchanged with microwave-frequency mechanical excitations.
As can be seen in Eq.(\ref{eq:OMintswap}), the optomechanical swapping interaction is enhanced by the average photon number, $\lvert\bar{a}\rvert^2$, which is proportional to the laser pump power.
Thus to maximize the effective optomechanical coupling, the laser pump should be as strong as possible before detrimental effects, such as heating, set in.

Previous work \cite{Bochmann_2013,VainsencherOMCbeam,Peairs_2020} and the majority of attention in optomechanical crystals for transduction has been on one-dimensional nanobeam designs \cite{Forsch_2019,Mirhosseini_2020,Jiang_2020,Arnold_2020,Stockill2021,Honl_2022}.
These structures exhibit strong optomechanical coupling, $g_\mathrm{om}/(2\,\pi) \sim 1\,\mathrm{MHz}$, and high mechanical and optical quality factors but are plagued by poor heat dissipation due to their one-dimensional nature limiting thermal conductivity \cite{MacCabe_2019}.
Recently, a quasi-two-dimensional optomechanical cavity design consisting of C-shape holes within the waveguide of a snowflake hole crystal lattice was demonstrated in silicon \cite{Ren_2020}.
This device showed improved heat management whilst maintaining good optomechanical coupling and quality factors, however, targeting $194\,\mathrm{THz}$ optics, operates at a mechanical frequency near $10\,\mathrm{GHz}$ --- somewhat above the optimal frequency for popular superconducting qubit archetypes.
Here, we adapt the design to gallium arsenide (GaAs), having half the acoustic wave speed of silicon, to produce a two-dimensional optomechanical resonator at a superconducting qubit compatible frequency $\sim 5\,\mathrm{GHz}$. Furthermore, the inherent piezoelectricity in gallium arsenide allows for future direct coupling to an electrical circuit.

\section{Design}

\begin{figure}
\includegraphics[width=\textwidth]{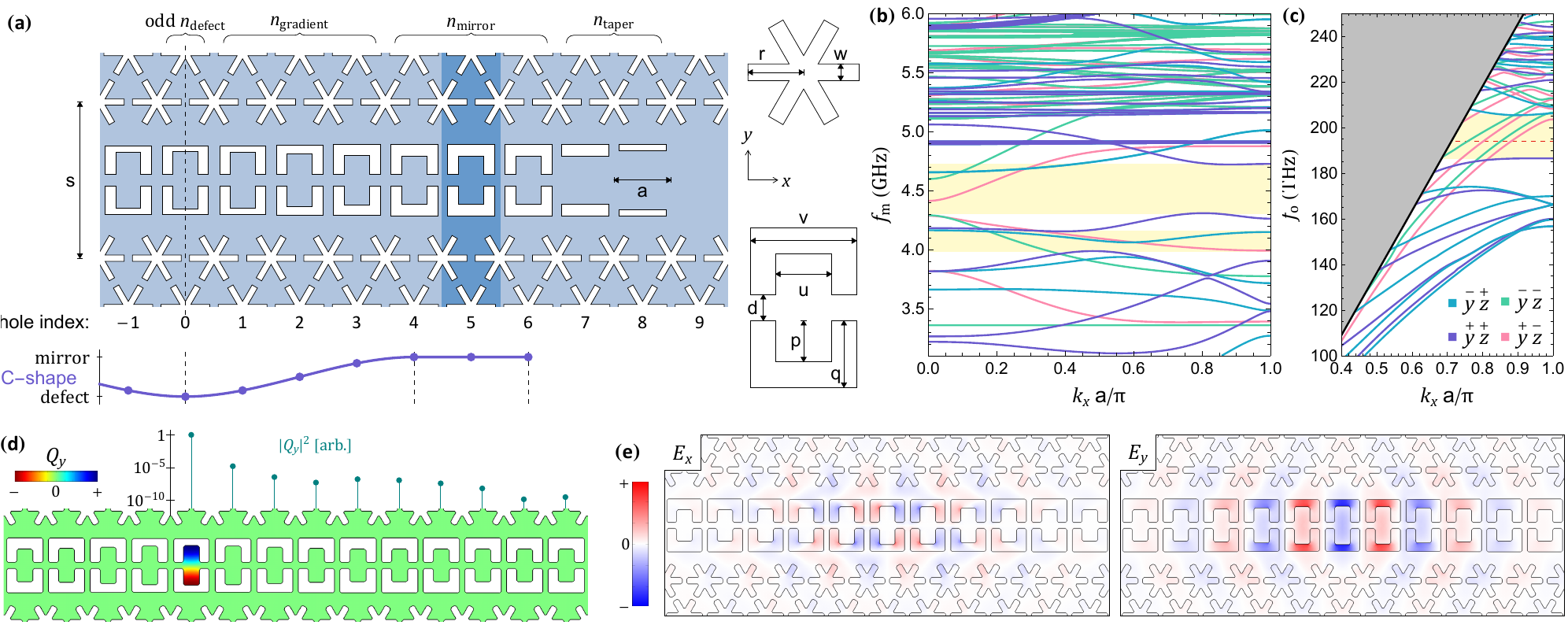}
\caption{
(a) Layout and geometric parameterization of the optomechanical crystal resonator. The crystal (mirror) unit cell is highlighted.
Parameter values in the gradient section between the defect and crystal unit cells are scaled with a $\mathrm{SmoothStep}_1$ function plotted by the purple line below.
(b) Acoustic dispersion relations from finite element model simulations of the mirror unit cell. Band gaps for $\overset{+}{y}\,\overset{+}{z}$ symmetric modes are highlighted by yellow bands.
(c) Electromagnetic dispersion relations from finite element model simulations of the mirror unit cell. Band gaps for $\overset{+}{z}$ symmetric modes are highlighted in yellow and cover our infra-red frequency of interest, $194\,\mathrm{THz}$, marked with a dashed red line.
(d) Simulation of the displacement field, $Q_y$, for the paddle breathing mode at $4.63\,\mathrm{GHz}$.
Displacements in paddles neighbouring the defect are too small to make out but are plotted above on a log scale.
(e) Simulations of the electric field for an electromagnetic resonance at $197\,\mathrm{THz}$.
Specific dimensions used for simulations are given in the supplementary materials \cite{supp}.
\label{fig:1}}
\end{figure}

To describe the symmetries present in the optomechanical crystal design we use the notation $\overset{\pm}{x}$ to denote a mirror ($+$) or anti-mirror ($-$) symmetry through the $x=0$ plane, such that $P_{\vec{x}} \cdot F[P_{\vec{x}} \cdot \vec{r}] = \pm F[\vec{r}]$ for some tensor field $F$, where $P$ is the Householder reflection transform.
The main two-dimensional optomechanical crystal is made from snowflake shaped holes in a periodic lattice that exhibits $\pi/3$ rotational symmetry and mirror symmetries (wallpaper group p6m).
This provides a complete acoustic band gap and $z$ mirror symmetric ($\overset{+}{z}$) electromagnetic band gap \cite{supp}.
Removing a row of snowflakes, and with additional stretching, creates a waveguide though the two-dimensional optomechanical crystal.
C-shape holes along this waveguide in a `vertebrae' pattern form a one-dimensional optomechanical crystal that can be gradually transitioned to create a defect cavity.
The layout and paramterization of these holes to form the resonator is given in Fig.~\ref{fig:1}(a).

C-shape hole dimensions in the gradient region between the mirror and defect are scaled according to the differentiable $\mathrm{SmoothStep}_1$ function, plotted in Fig.~\ref{fig:1}(a),
with definition $\mathrm{SmoothStep}_n:[0,1]\rightarrow[0,1]$ by $x\mapsto x^{n+1}\,\sum_{k=0}^n\,(\begin{smallmatrix}n+k\\k\end{smallmatrix})\,(\begin{smallmatrix}2\,n+1\\n-k\end{smallmatrix})\,(-x)^k$, where $(\begin{smallmatrix}n\\k\end{smallmatrix})$ are binomial coefficients.
This is the lowest order polynomial between constant zero and one that is $n$-differentiable.

Here, we focus on $n_\text{defect}=1$ single paddle resonances.
Although $n_\text{defect}>1$ resonators provide better optomechanical couplings in simulation, they are prone to fabrication imperfections, splitting the resonance into separate modes. Simulations showed that a $+1\,\mathrm{nm}$ change in paddle length ($\mathsf{p}$) shifted the individual resonance by $-17\,\mathrm{MHz}$.

Starting with C-shape hole parameters used in silicon \cite{Ren_2020, Kersul2022}, finite element model simulations \cite{comsol} were carried out in gallium arsenide and iterated upon to find a set of dimensions that produced the desired band gaps.
Dispersion relations for such parameters over a vertebrae unit cell are given in Fig.~\ref{fig:1}(b,c).
The acoustic resonance displacement field must be mirror-symmetric about every geometric mirror plane in order to have non-vanishing optomechanical coupling \cite{PoveyThesis}. For the vertebrae resonator design, this is the paddle breathing mode with mirror symmetries about each cardinal plane through the center ($\overset{+}{x}\,\overset{+}{y}\,\overset{+}{z}$) as depicted in Fig.~\ref{fig:1}(d).
By changing the paddle length ($\mathsf{p}$) between the mirror cell and defect cell, we can move the breathing mode from outside the band gap to within.

Eigenvalue simulations of the full vertebrae resonator are shown in Fig.~\ref{fig:1}(d,e).
In order to accommodate future acoustic piezoelectric wave propagation, the waveguide is aligned along the $[110]$ GaAs crystal axis \cite{PoveyThesis}.
The optomechanical coupling, calculated from displacement and electric fields, due to the moving boundary and photo-elastic effect, is $g_\mathrm{om}/(2\,\pi) = 599\,\mathrm{kHz}$.

\section{Fabrication}

\begin{figure}
\includegraphics[width=\textwidth]{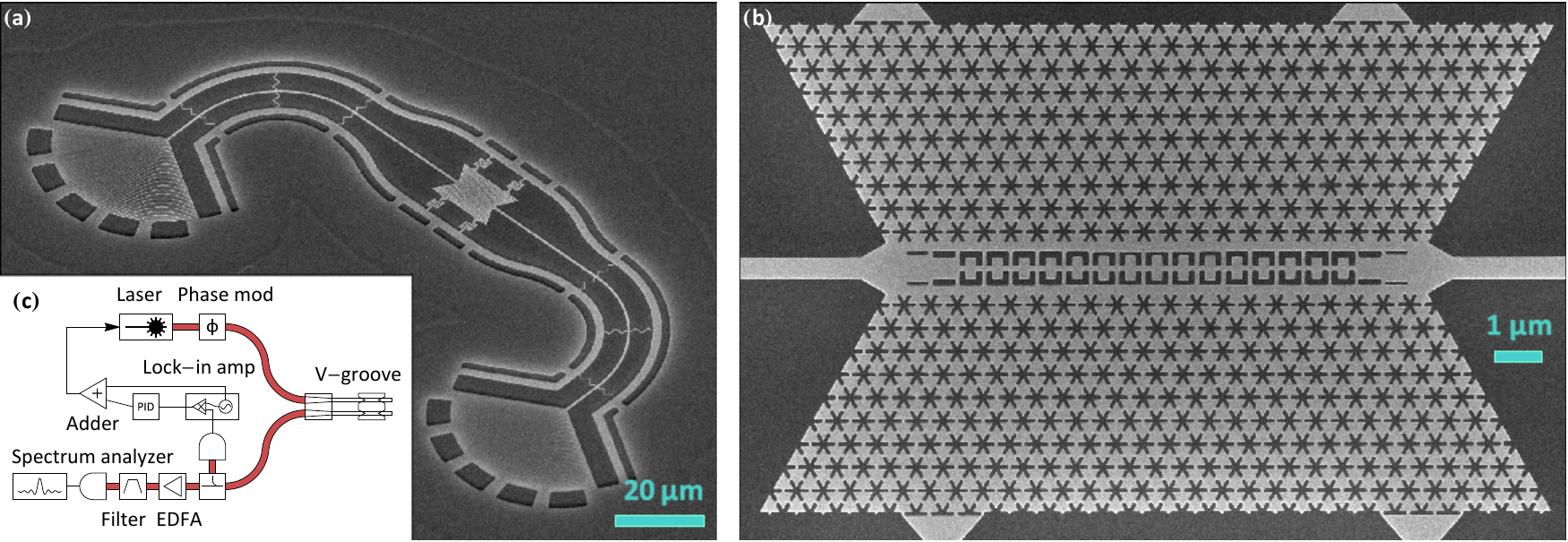}
\caption{
(a) Scanning electron microscope image, taken at $45^\circ$, of a device similar to the one measured here. Two large suspended grating coupler pads deflect light in and out of plane to couple to overhead optical fibers. The optomechanical crystal plate is suspended in the middle with large serpentine tethers to relieve tension during the release process.
(b) Magnified image of the optomechanical plate from (a), taken top down, showing the vertebrae resonator. Proximity effect correction during electron beam lithography is essential for uniform snowflakes.
(c) Simplified diagram of the optical measurement setup showing only elements discussed in the text. The phase modulator is driven by a microwave frequency signal generator and provides a calibration of power measured by the spectrum analyzer. The V-groove assembly holds fiber optics above the device. A portion of the output signal is used in a dither locking system, utilizing a lock-in amplifier and proportional-integral-derivative (PID) controller, that adjusts the laser output wavelength. The detection chain consists of an erbium doped fiber amplifier (EDFA) followed by a filter to cut out amplified spontaneous emission, and then a microwave-frequency bandwidth photodetector.
\label{fig:2}}
\end{figure}

Devices are fabricated on heterostructure wafers consisting of $250\,\mathrm{nm}$ $\mathrm{Ga}\mathrm{As}$, on a $1\,\upmu\mathrm{m}$ $\mathrm{Al}_{0.9}\mathrm{Ga}_{0.1}\mathrm{As}$ sacrificial layer, on $635\,\upmu\mathrm{m}$ bulk $(100)$ $\mathrm{Ga}\mathrm{As}$.
The relatively low suspension height provided by the sacrificial layer ($\mathsf{t}_\text{gap} < \lambda_\mathrm{o}$) leads to optical fields leaking into the bulk, limiting optical quality factors and grating coupler performance.
Patterning is performed with electron beam lithography using $14\%$ by weight hydrogen silsesquioxane in methyl isobutyl ketone resist and a chlorine, argon inductively coupled plasma etch.
Proximity effect correction is vital when patterning nanometer scale features over hundreds of square microns on a heavy substrate with significant electron back scatter, details of our point spread function used are provided in the supplementary materials \cite{supp}.
To account for uncertainty in precise slab thickness, produced hole sizes, and simulations, devices are fabricated in series of varying global scale factor, typically in steps of $1\%$.
Parameter dimensions for the pattern used to write the device presented here are listed in the supplementary materials \cite{supp}.

To couple light to the device, two suspended grating coupler pads are used to deflect light in and out of plane to optical fibers positioned above the device.
The resonator is then interposed along a connecting suspended waveguide as depicted in Fig.~\ref{fig:2}(a,b).
Although less efficient than tapered-beam edge coupling, this approach allowed up to 100 devices to be fabricated and tested on a single $1\,\mathrm{cm}^2$ chip.

\section{Measurement}

\begin{figure}
\includegraphics[width=0.45\textwidth]{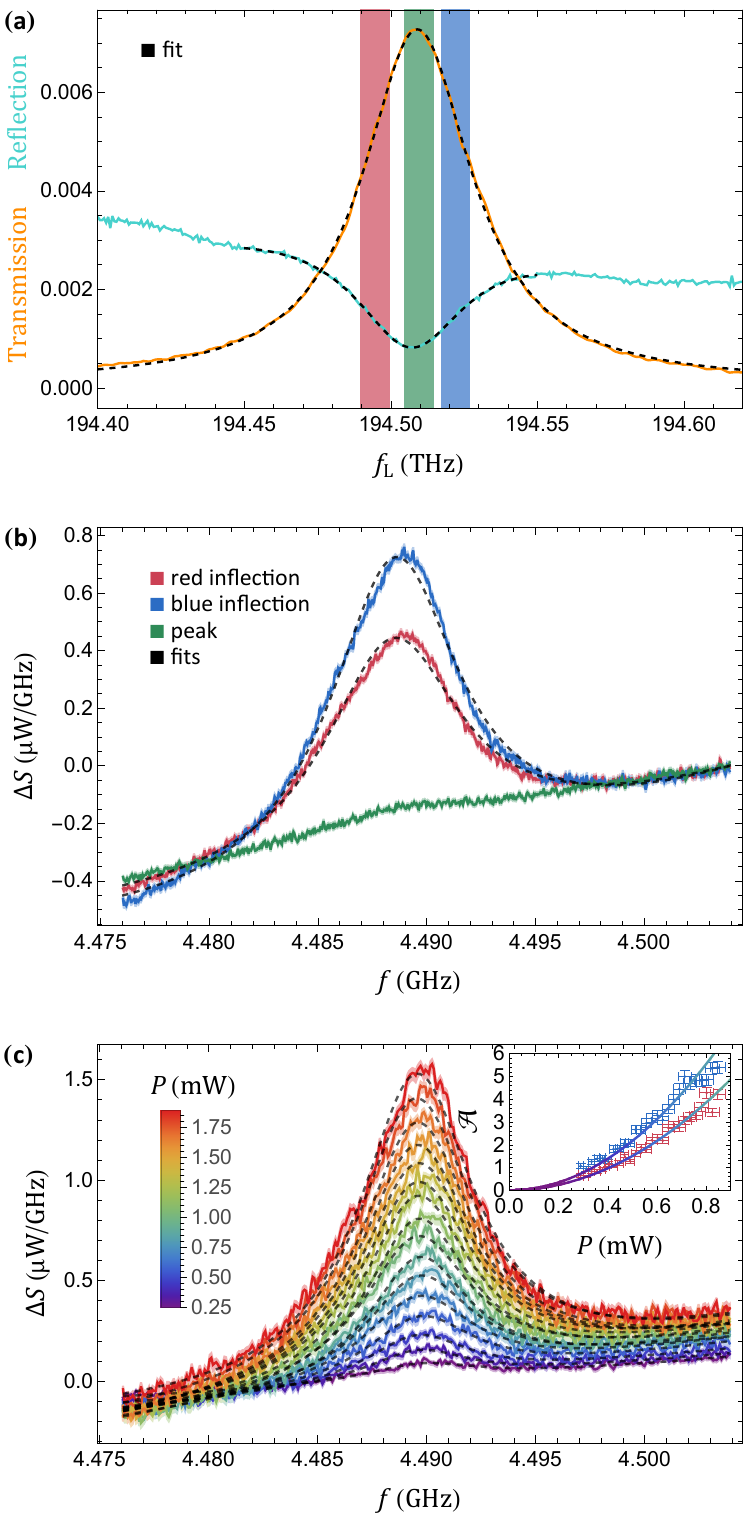}
\caption{
(a) Optical resonance transmission (orange) and reflection (cyan) with colour-coded probing regions covered by the dither lock.
(b) Thermal mechanical spectrum imprinted on the optical signal when probed at the optical resonance and inflection points (colour coded accordingly). Each trace is the average of 100 scans with a 100 point moving average, statistical uncertainty of $\pm1$ standard deviation is depicted in light shading about the trace.
(c) Scans of the thermal mechanical spectrum with different input powers, referenced to the V-groove assembly. Inset shows fit spectrum amplitude $\mathcal{A}$ as a function of input power over our photodetector's linear operating regime, with $P^2$ fit curves.
\label{fig:3}}
\end{figure}

Optical measurements are carried out at room temperature using an infra-red fiber optic setup and tunable laser.
A simplified diagram including only major elements is shown in Fig.~\ref{fig:2}(c).
To compensate for slow thermal drifts of the optical resonance frequency and overall transmission, a second-harmonic dither locking system is used to track the resonance inflection points.
A diagram of the complete setup, and details on the dither locking, are included in the supplementary material \cite{supp}.

Transmission through the optical resonance is given by
\begin{equation}
\lvert S_{2\,1}(\omega)\rvert^2 = \frac{\left(\gamma_\mathrm{o}^\mathrm{wg}\right)^2}{\left(\omega-\omega_\mathrm{o}\right)^2 + \left(\frac{\gamma_\mathrm{o}}{2}\right)^2} \nonumber\;,
\end{equation}
with $\gamma_\mathrm{o} = \gamma_\mathrm{o}^0 + 2\,\gamma_\mathrm{o}^\mathrm{wg}$,
where $\gamma_\mathrm{o}^0$ is intrinsic power loss rate and $\gamma_\mathrm{o}^\mathrm{wg}$ is power loss rate to each waveguide port.
The loaded quality factor is defined as $Q_\mathrm{o}^\mathrm{L} = \omega_\mathrm{o}/\gamma_\mathrm{o}$, and the internal quality factor is $Q_\mathrm{o}^0 = \omega_\mathrm{o}/\gamma_\mathrm{o}^0$.
A captured transmission profile is given in Fig.~\ref{fig:3}(a).

At room temperature, the mechanical resonance has a thermal population $\bar{n}_\mathrm{m} \approx \frac{k_\mathrm{B}\,T}{\hbar\,\omega_\mathrm{m}} = 1371\pm14$, which through the optomechanical interaction will modulate the optical resonance at the mechanical resonance frequency. Probing the side of the optical resonance, we are sensitive to the resulting power fluctuation using a high bandwidth photodetector, and measuring with a spectrum analyzer can observe the mechanical thermal spectrum.

Conveniently, the transmission factor of this optomechanical spectrum is identical to the transmission factor of a weak phase-modulated signal \cite{Gorodetsky_2010}, allowing us to make a calibrated measurement of the optomechanical coupling.
At laser frequency $\omega_\mathrm{L}$, the phase modulated input is $\exp\bigl[-\mathrm{i}\,\omega_\mathrm{L}\,t+\mathrm{i}\,A_\Phi\,\sin[\omega_\Phi\,t]\bigr]$.
For hot optomechanics, $\bar{n}_\mathrm{m} \gg 1$, and weak phase modulation, $A_\Phi \ll 1$, our power spectral density is, to leading order in $A_\Phi$, \cite{PoveyThesis}
\begin{equation}
S_{XX}(\omega>0) \approx \mathcal{T}(\omega_\mathrm{L};\omega) \, \Bigl(
{g_\mathrm{om}}^2\,\frac{\bar{n}_\mathrm{m}\,\gamma_\mathrm{m}}{\left(\omega-\omega_\mathrm{m}\right)^2+\left(\frac{\gamma_\mathrm{m}}{2}\right)^2}
+\frac{{A_\Phi}^2\,{\omega_\Phi}^2}{4}\,2\,\pi\,\delta(\omega-\omega_\Phi)
\Bigr) \nonumber\;,
\end{equation}
where $\gamma_\mathrm{m}=\omega_\mathrm{m}/Q_\mathrm{m}$ is the mechanical mode power loss rate,
and $\mathcal{T}(\omega_\mathrm{L};\omega)$ is a complicated function of the laser frequency and spectrum frequency that includes transmission through the optical cavity and all efficiencies.
Using a spectrum analyzer with window function $w(f_\mathrm{SA},\mathrm{RBW};f)$, where $f_\mathrm{SA}$ is the measurement frequency and $\mathrm{RBW}$ is the resolution bandwidth,
with step size $\Delta f_\mathrm{SA} \ll \mathrm{RBW} \ll \gamma_\mathrm{m}/(2\,\pi)$, the measured power is
\begin{equation}
P(\omega_\mathrm{SA}) = \mathcal{T}(\omega_\mathrm{L};\omega_\mathrm{SA})\,\Bigl(
\mathrm{RBW}\,2\,{g_\mathrm{om}}^2\,\frac{\bar{n}_\mathrm{m}\,\gamma_\mathrm{m}}{\left(\omega_\mathrm{SA}-\omega_\mathrm{m}\right)^2+\left(\frac{\gamma_\mathrm{m}}{2}\right)^2}
+ \frac{{A_\Phi}^2\,{\omega_\Phi}^2}{2}\,w(\omega_\mathrm{SA},\mathrm{RBW};\omega_\Phi)
+ \text{bg}
\Bigr) \nonumber\;.
\end{equation}
A broad background, denoted by $\mathrm{bg}$, can be removed with a low-order polynomial fit.
If $\omega_\Phi \approx \omega_\mathrm{m}$, then $\mathcal{T}(\omega_L;\omega_\mathrm{SA})$ is approximately constant.
Fitting the window function over the phase modulated signal (with known amplitude $A_\Phi$) allows us to determine $\mathcal{T}$, then fitting the thermal spectrum gives us $g_\mathrm{om}$.
For fast scans with no averaging, there is a dither imprint that can be included in the fit or ignored.

In practice, a fit over the thermal spectrum gives us $\omega_\mathrm{m}$, $\gamma_\mathrm{m}$, and the parameter $\mathcal{A} = 2\,\mathcal{T}\,{g_\mathrm{om}}^2\,\bar{n}_\mathrm{m}\,\gamma_\mathrm{m}$, whilst a window fit over the phase modulation signal gives us $P_{f_\Phi} = \mathcal{T}\,{A_\Phi}^2\,{\omega_\Phi}^2/2$, such that
\begin{equation}
g_\mathrm{om} = \frac{\omega_\Phi\,A_\Phi}{\sqrt{k_\mathrm{B}\,T/\hbar}}\,\sqrt{\frac{\omega_\mathrm{m}\,\mathcal{A}}{4\,\gamma_\mathrm{m}\,P_{f_\Phi}}} \nonumber\;.
\end{equation}
Spectrum scans are fit individually with each producing $\sqrt{\frac{\omega_\mathrm{m}\,\mathcal{A}}{4\,\gamma_\mathrm{m}\,P_{f_\Phi}}}$, and these are combined together to give a factor that contains all the statistical uncertainty.
Temperature, $T$, and phase modulation strength, $A_\Phi$, are included with systematic uncertainties.

The inset of Fig.~\ref{fig:3}(a) shows the optical resonance transmission and reflection, fits give $f_\mathrm{o}=194.5\,\mathrm{THz}$ with loaded quality factor $Q_\mathrm{o}^\mathrm{L}\approx4300$ and internal quality factor $Q_\mathrm{o}^\mathrm{0}\approx8600$.
Using a first-harmonic dither lock to the resonance peak (shown in green) there is no optomechanical spectrum visible, as expected.
With a second-harmonic dither lock to either of the resonance inflection points (blue and red), the thermal optomechanical spectrum can be seen at $f_\mathrm{m}=4.488\,\mathrm{GHz}$ with quality factor $Q_\mathrm{m} \approx 600$.
Fig.~\ref{fig:3}(b) shows the optomechanical spectrum averaged over 100 scans.

Optomechanical couplings are calculated for each individual scan by fitting the thermal spectrum and phase modulate signal.
The weighted mean from 100 scans at both the blue and red inflection points yielded $g_\mathrm{om}/(2\,\pi) = (649 \pm 8)\,\mathrm{kHz}$.

Scans made at varied laser probe powers are shown in Fig.~\ref{fig:3}(c).
The current output of our photodetector is proportional to the laser power received, giving the power of the measured signal proportional to the square of laser power.
This scaling of the thermal power spectrum amplitude, $\mathcal{A}$, is shown in the inset.

\section{Discussion}
We have demonstrated a quasi-two-dimensional optomechanical resonator in gallium arsenide, using the vertebrae design, with an optical resonance in the infra-red C band and mechanical resonance at $\sim4.5\,\mathrm{GHz}$ compatible with transmon superconducting qubits.
The optomechanical coupling of $\sim 650\,\mathrm{kHz}$ is slightly less than the best demonstrated couplings of $\sim 1\,\mathrm{MHz}$ \cite{Forsch_2019,Ren_2020}, however there is significant room for improvement in design parameter optimization, fabrication, and cryogenic operation.
Switching to a $3\,\upmu\mathrm{m}$ $\mathrm{Al}_{0.7}\mathrm{Ga}_{0.3}\mathrm{As}$ sacrificial layer should provide easily obtainable gains \cite{Forsch_2019}.
Transitioning to edge coupling in reflection will open up the other side of the vertebrae resonator to electromechanical coupling.

\section{Acknowledgements}
The authors would like to thank Peter Duda for helpful discussions.
This work was supported by the LPS/ARO award W911NF-23-1-0077, NSF QLCI for HQAN (award 2016136), by the Air Force Office of Scientific Research, and in part based on work supported by the U.S. Department of Energy Office of Science National Quantum Information Science Research Centers, and by UChicago’s MRSEC (NSF award DMR-2011854). 
This work made use of the Pritzker Nanofabrication Facility, part of the Pritzker School of Molecular Engineering at the University of Chicago, which receives support from Soft and Hybrid Nanotechnology Experimental (SHyNE) Resource (NSF ECCS-2025633), a node of the National Science Foundation’s National Nanotechnology Coordinated Infrastructure.

\clearpage
\bibliography{jabref.bib}

\clearpage
\appendix

\section{Snowflake crystal dispersion}
\begin{figure}[tb]
\centering
\includegraphics[width=\textwidth]{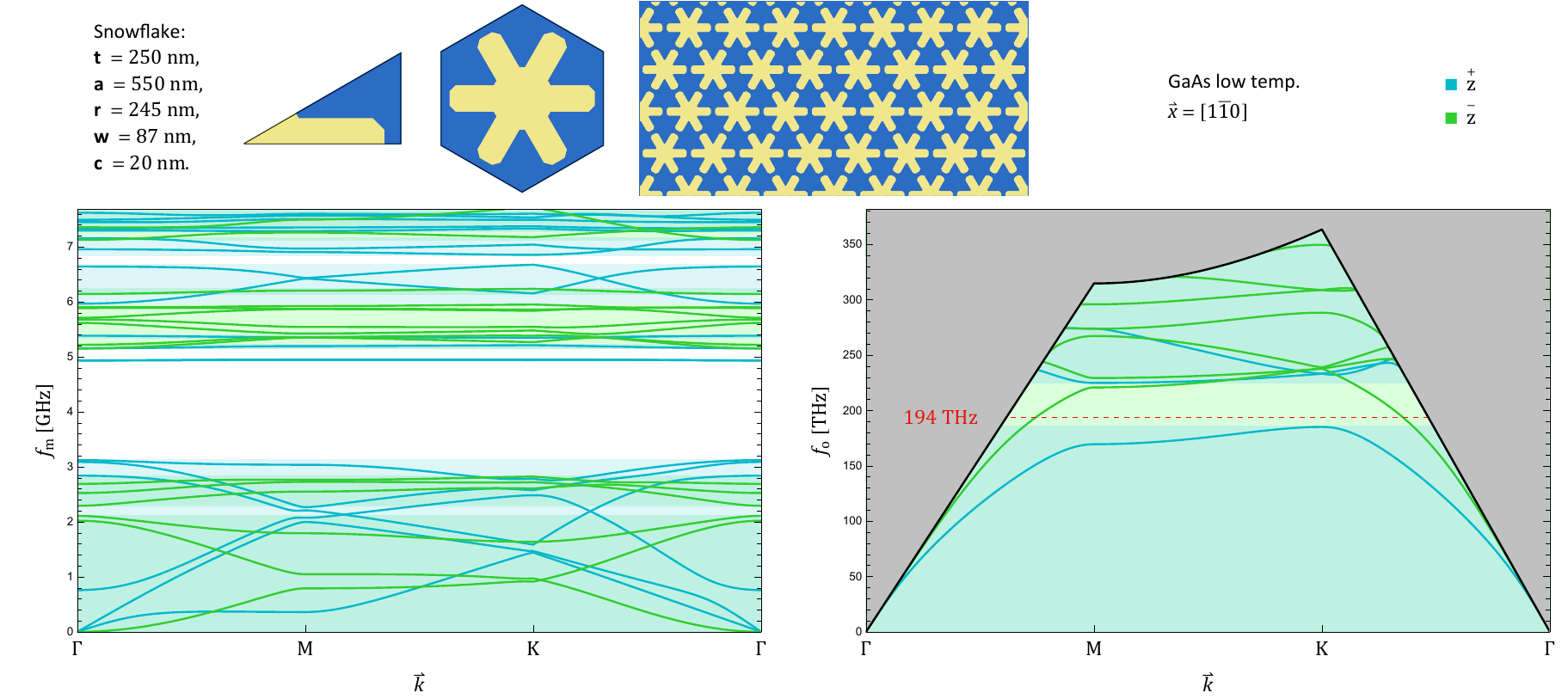}
\caption{Snowflake in GaAs, (left) acoustic and (right) electromagnetic dispersion plots from finite element simulations.
\label{fig:SnowflakeDisps}}
\end{figure}
Plots of the two-dimensional snowflake optomechanical crystal dispersion relations are given in Fig.~\ref{fig:SnowflakeDisps}.
Snowflake parameters are defined in Fig.~1.

\section{Parameter specifications}
\begin{figure}[tb]
\centering
\includegraphics[width=0.75\textwidth]{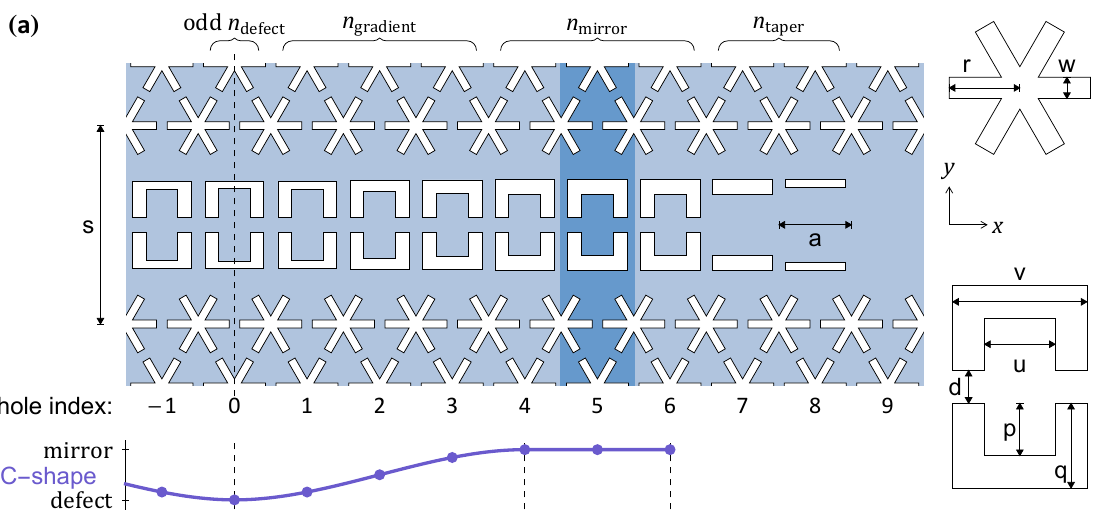}
\caption{Fig~1(a) showing the vertebrae resonator dimension specifications. \label{fig:vertebraedims}}
\end{figure}
\begin{table}
\begin{tabular}{l|r|r}
Parameter & Simulation & Fabrication \\
\hline
$\mathsf{t}$ & $250\,\mathrm{nm}$ & $\approx250\,\mathrm{nm}$ \\
$\mathsf{a}$ & $550\,\mathrm{nm}$ & $572\,\mathrm{nm}$ \\
$\mathsf{r}$ & $245\,\mathrm{nm}$ & $245\,\mathrm{nm}$ \\
$\mathsf{w}$ & $87\,\mathrm{nm}$ & $62\,\mathrm{nm}$ \\
$\mathsf{q}_\text{mirror}$ & $320\,\mathrm{nm}$ & $299\,\mathrm{nm}$ \\
$\mathsf{v}_\text{mirror}$ & $480\,\mathrm{nm}$ & $472\,\mathrm{nm}$ \\
$\mathsf{p}_\text{mirror}$ & $175\,\mathrm{nm}$ & $182\,\mathrm{nm}$ \\
$\mathsf{u}_\text{mirror}$ & $210\,\mathrm{nm}$ & $252\,\mathrm{nm}$ \\
$\mathsf{q}_\text{defect}$ & $310\,\mathrm{nm}$ & $285\,\mathrm{nm}$ \\
$\mathsf{v}_\text{defect}$ & $470\,\mathrm{nm}$ & $453\,\mathrm{nm}$ \\
$\mathsf{p}_\text{defect}$ & $220\,\mathrm{nm}$ & $228\,\mathrm{nm}$ \\
$\mathsf{u}_\text{defect}$ & $210\,\mathrm{nm}$ & $249\,\mathrm{nm}$ \\
$\mathsf{d}$ & $80\,\mathrm{nm}$ & $115\,\mathrm{nm}$ \\
$\mathsf{s}$ & $1503\,\mathrm{nm}$ & $1563\,\mathrm{nm}$ \\
$n_\text{defect}$ & $1$ & $1$ \\
$n_\text{gradient}$ & $3$ & $3$ \\
$n_\text{mirror}$ & $6$ & $3$ \\
$n_\text{taper}$ & $0$ & $2$ \\
\end{tabular}
\caption{Parameters used for finite element model simulations and what was eventually used in the electron beam lithography pattern.
\label{tab:dims}}
\end{table}
Dimensional parameters, defined in Fig.~1(a) (reproduced in Fig.~\ref{fig:vertebraedims}) with the addition of $\mathsf{t}$ for slab thickness, used for given simulations (Fig.~1) and the electron beam lithography pattern of the measured device (Fig.~3) are listed in Tab.~\ref{tab:dims}.
Snowflake and C-shape holes in simulations include a chamfer of $\mathsf{c}=20\,\mathrm{nm}$ to account for imperfect corners in fabrication.

GaAs is aligned such that $[110] \parallel \vec{x}$ along the waveguide direction. This is chosen to accommodate future electromechanical elements --- symmetries in the displacement and voltage fields restrict piezoelectric wave propagation in GaAs to along $[110]$.

\begin{figure}[tb]
\centering
\includegraphics[width=0.75\textwidth]{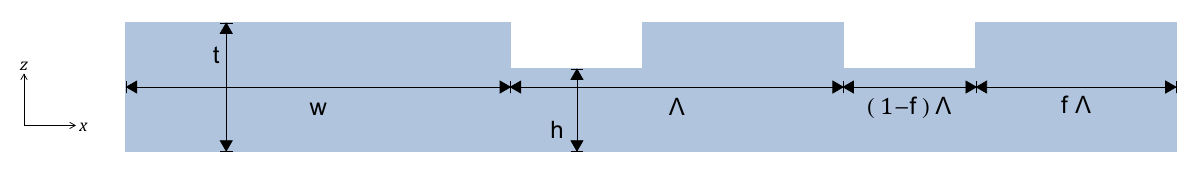}
\caption{Diagram of grating coupler dimensions. \label{fig:gratingdims}}
\end{figure}
\begin{table}
\begin{tabular}{l|r}
Parameter & Fabrication \\
\hline
$\mathsf{t}$ & $\approx 250\,\mathrm{nm}$ \\
$\mathsf{h}$ & $\approx 150\,\mathrm{nm}$ \\
$n_\mathrm{grates}$ & $40$ \\
$\mathsf{\Lambda}$ & $632\,\mathrm{nm}$ \\
$\mathsf{f}$ & $0.58$ \\
$b$ & $-0.05$ \\
$\mathsf{w}$ & $1\,\upmu\mathrm{m}$
\end{tabular}
\caption{Grating coupler parameters used in photolithography.
\label{tab:grating}}
\end{table}
The grating couplers, with dimensions defined in Fig.~\ref{fig:gratingdims}, have parameters given in Tab.~\ref{tab:grating}, with curves following
\begin{equation}
a = b\,x+\sqrt{x^2+y^2} \label{eq:curvefunc} \;,
\end{equation}
about focus $(0,0)$, where $a$ is the $y$-intercept distance.

Tethers in a zigzag pattern to hold up the suspended beam waveguide are kept thin ($125\,\mathrm{nm}$) such as not to support optical modes.
Larger serpentine tethers hold up the main optomechanical crystal slab and provide some give during the release process.
Windows around the open area mitigate possible damage from large cracks that form during the release.
A gentler release (with lower aluminium content sacrificial layer) could allow for a more direct connection.

\begin{table}
\begin{tabular}{l|r}
Parameter & Simulation \\
\hline
$\rho$ & $5317\,\mathrm{kg}\,\mathrm{m}^{-3}$ \\
$\tilde{c}^{1\,1}$ & $118.41\,\mathrm{GPa}$ \\
$\tilde{c}^{1\,2}$ & $53.78\,\mathrm{GPa}$ \\
$\tilde{c}^{4\,4}$ & $59.12\,\mathrm{GPa}$ \\
$\tilde{e}^{1\,4}$ & $-0.16\,\mathrm{C}\,\mathrm{m}^{-2}$ \\
$\tilde{p}^{1\,1}$ & $-0.165$ \\
$\tilde{p}^{1\,2}$ & $-0.140$ \\
$\tilde{p}^{4\,4}$ & $-0.072$ \\
$\varepsilon_\mathrm{r}$ & $11.361$
\end{tabular}
\caption{GaAs material properties used in finite element modelling.
\label{tab:material}}
\end{table}
Gallium arsenide material properties are listed in Tab.~\ref{tab:material}.
$\rho$ is density, $c$ is stiffness, $p$ is photo-elastic, and $\varepsilon_\mathrm{r}$ is relative permittivity.

\section{Optomechanical coupling calculation}
The optomechanical coupling was calculated from simulations according to
\begin{equation}
g_\mathrm{om} = \left\lvert g_{\substack{\mathrm{om}\\\mathrm{MB}}} + g_{\substack{\mathrm{om}\\\mathrm{PE}}}\right\rvert \nonumber\;.
\end{equation}
The moving boundary contribution is
\begin{multline}
g_{\substack{\mathrm{om}\\\mathrm{MB}}} = -\frac{\omega_\mathrm{o}}{2}\sqrt{\frac{\hbar}{2\,\omega_\mathrm{m}}} \\ 
\frac{\int_{\partial S} Q_k(\vec{r})\,n_k(\vec{r})\,\left(
E^\parallel_i(\vec{r})^\ast\,\left( (\varepsilon_S)_{i\,j}-(\varepsilon_0)_{i\,j}\right)\,E^\parallel_j(\vec{r}) - D^\perp_i(\vec{r})^\ast\,\Bigl( {(\varepsilon_S)^{-1}}_{i\,j}-{(\varepsilon_0)^{-1}}_{i\,j}\Bigr)\,D^\perp_j(\vec{r})
\right)\,\mathrm{d}^2\vec{r}}{\sqrt{\int_S Q_i(\vec{r})^\ast\,\rho(\vec{r})\,Q_i(\vec{r})\,\mathrm{d}^3\vec{r}}\quad\int_V E_i(\vec{r})^\ast\,\varepsilon_{i\,j}(\vec{r})\,E_j(\vec{r})\,\mathrm{d}^3\vec{r}} \nonumber\;,
\end{multline}
where $\vec{Q}$ is the complex harmonic displacement field, $\vec{E}$ and $\vec{D}$ are the complex harmonic electric and electric displacement fields, $\rho$ is density, $\varepsilon$ is permittivity, $V$ is all space, and $S$ is the solid with boundary $\partial S$ and normal $\vec{n}$.
The photoelectric contribution is
\begin{equation}
g_{\substack{\mathrm{om}\\\mathrm{PE}}} = \frac{\omega_\mathrm{o}}{2}\sqrt{\frac{\hbar}{2\,\omega_\mathrm{m}}}
\frac{\frac{1}{\varepsilon_0}\,\int_S E_i(\vec{r})^\ast\,(\varepsilon_S)_{i\,a}\,p_{a\,b\,c\,d}\,\epsilon_{c\,d}(\vec{r})\,(\varepsilon_S)_{b\,j}\,E_j(\vec{r})\,\mathrm{d}^3\vec{r}}{\sqrt{\int_S Q_i(\vec{r})^\ast\,\rho(\vec{r})\,Q_i(\vec{r})\,\mathrm{d}^3\vec{r}}\quad\int_V E_i(\vec{r})^\ast\,\varepsilon_{i\,j}(\vec{r})\,E_j(\vec{r})\,\mathrm{d}^3\vec{r}} \nonumber\;,
\end{equation}
where $\epsilon$ is strain, and $p$ is the photo-elastic tensor noting that in Voigt notation it transforms stress-like.

\section{Proximity effect correction}
Due to the greater atomic weight of gallium and arsenide (compared to silicon, for example), electron back scatter during the pattern writing process leads to significant proximity dependent over-dosing.
This is particularly problematic for fashioning uniform nanoscale features over large areas, as is needed for a two-dimensional optomechanical crystal.
To combat this we employ proximity effect corrections using a point spread function developed over many fabrication cycles.

Our point spread function is modelled using Gaussian-Pearson VII product terms of the form
\begin{equation}
f_\mathrm{GP}[\sigma,\gamma,\nu;r]=\frac{\exp\bigl[-\gamma^2/\sigma^2\bigr]}{\pi\,\gamma^2\,\mathrm{E}_\nu\bigl[\gamma^2/\sigma^2\bigr]}\,\exp\bigl[-r^2/\sigma^2\bigr]\,\left(\frac{r^2}{\gamma^2}+1\right)^{-\nu} \;,
\end{equation}
where $E$ is the exponential integral function.
This form is able to include a basic Gaussian scattering radius, $\sigma$, and diffusion-like process with range $\gamma$ and exponent $\nu$.

For fabricating snowflake crystal patterns, the normalized point spread function was
\begin{align}
\mathrm{PSF}[r] =&\; 0.143885\,f_\mathrm{GP}[685\,\mathrm{nm},1\,\mathrm{nm},1;r] \nonumber\\
&\; + 0.172662\,f_\mathrm{GP}[5\,\mathrm{nm},0,0;r] \nonumber\\
&\; + 0.683453\,f_\mathrm{GP}[13\,\upmu\mathrm{m},0,0;r] \nonumber\;,
\end{align}
implemented with a short range cut-off of $0.1\,\upmu\mathrm{m}$ .

\section{Optical measurement setup}
\begin{figure}[p]
\centering
\includegraphics[width=\textwidth]{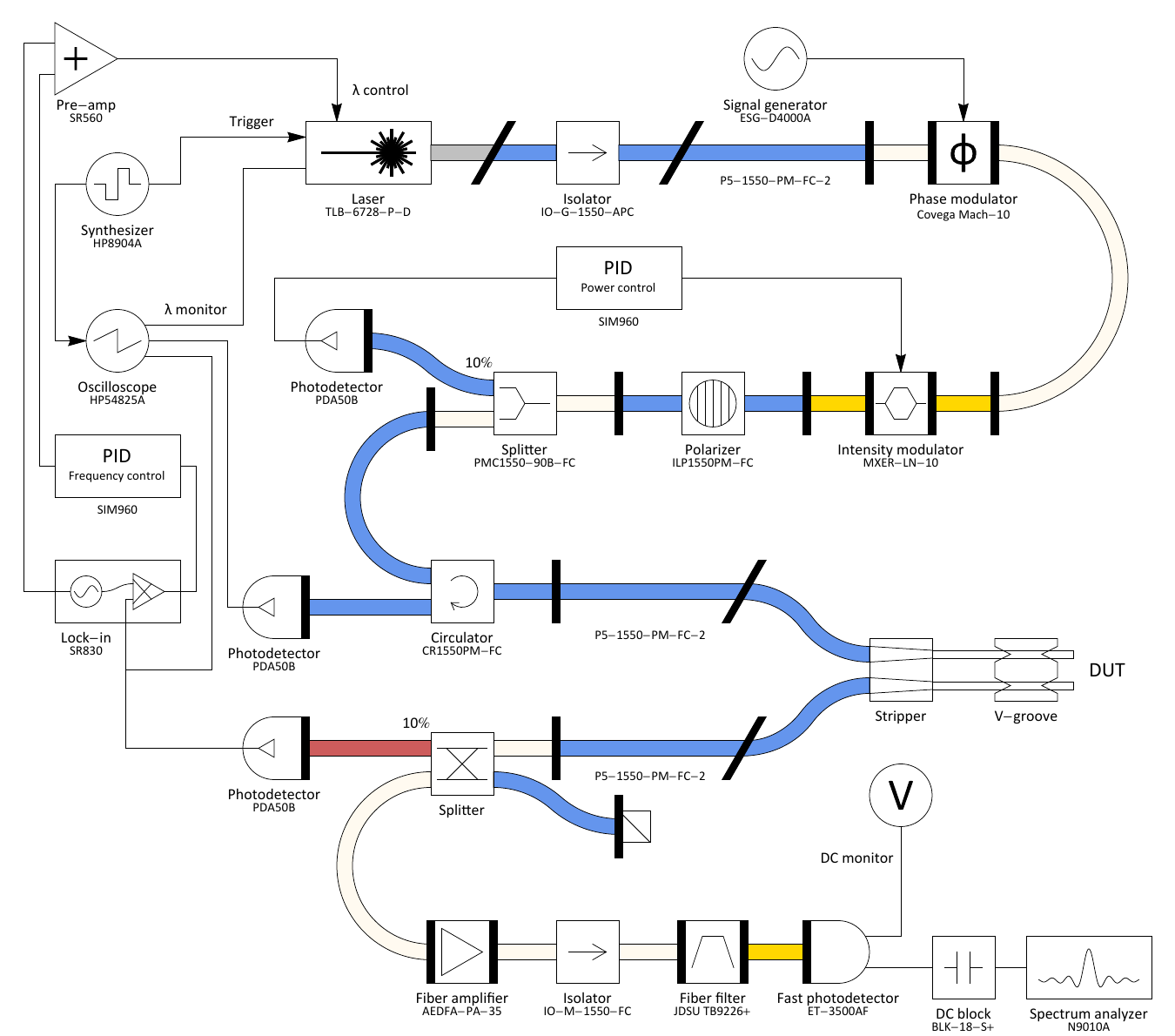}
\caption{Full diagram of the optical measurement setup. Thick vertical lines indicate flat physical contact (PC) connections, whilst thick angled lines indicate angled physical connections (APC). Fiber optic colours mimic their real-life colour, lengths are not to scale. \label{fig:opticssetup}}
\end{figure}
The measurement setup is a C band infra-red fiber optic based system with free space coupling to a chip in transmission and a high bandwidth photodetector that can resolve microwave frequencies. A complete diagram of the system is given in Fig.~\ref{fig:opticssetup}.
From the laser, the optic signal first goes through optional phase modulation, and then a variable attenuator for power control, and a polarizer. It is directed onto the chip by a V-groove clamp that holds the fiber in place above the chip at an angle of $8^\circ$ from normal.
This angle was optimized for silicon on insulator grating couplers of a previous experiment.
A second fiber is clamped $127\,\upmu\mathrm{m}$ away to pick up the transmitted signal. A splitter takes a small amount of the signal for frequency locking whilst the rest goes through an amplifier and filter before hitting the fast photodetector.

Simple side-locking to a transmission value is not adequate for tracking the optical resonance edge due to the prevalence of global efficiency drifts thought to be due to temperature fluctuations. To deal with both frequency and transmission drift, a dither lock feedback system is implemented. The laser wavelength is modulated at $f_\text{dither} = 100\,\mathrm{Hz}$ and a lock-in amplifier operating at the $n$'th harmonic measures the $n$'th derivative of the output transmission profile. A proportional-integral-derivative controller, with a set point of 0 for the aforementioned signal, shifts the laser center wavelength accordingly. Tracking the inflection point ($2$nd derivative equal to zero) is done using the second harmonic.

\end{document}